\documentclass[%
 aip,
 jcp,
 amsmath,amssymb,
 preprint,%
 longbibliography,%
]{revtex4-1}

\usepackage{graphicx}
\usepackage{dcolumn}
\usepackage{bm}

\usepackage[colorlinks=false,citecolor=blue,breaklinks=true]{hyperref}
\usepackage[utf8]{inputenc}
\usepackage[T1]{fontenc}
\usepackage{mathptmx}

\graphicspath{{images/}}
\usepackage[normalem]{ulem}

\usepackage[version=4]{mhchem}
\usepackage{siunitx}

\begin{document}


\title{ The Role of High-Order Electron Correlation Effects in a Model System for Non-valence Correlation-bound Anions}

\author{Shiv Upadhyay}
\affiliation{Department of Chemistry, University of Pittsburgh, Pittsburgh, Pennsylvania 15260, USA}
\author{Amanda Dumi}
\affiliation{Department of Chemistry, University of Pittsburgh, Pittsburgh, Pennsylvania 15260, USA}
\author{James Shee}
\affiliation{Department of Chemistry, University of California Berkeley, Berkeley, California, 94720, USA}
\author{Kenneth D. Jordan}
\email{jordan@pitt.edu}
\affiliation{Department of Chemistry, University of Pittsburgh, Pittsburgh, Pennsylvania 15260, USA}

\date{\today}

\begin{abstract}
    The diffusion Monte Carlo (DMC), auxiliary field quantum Monte Carlo (AFQMC), and equation-of-motion coupled cluster (EOM-CC) methods are used to calculate the electron binding energy (EBE) of the non-valence anion state of a model \ce{(H2O)4} cluster. Two geometries are considered, one at which the anion is unbound and the other at which it is bound in the Hartree-Fock (HF) approximation. It is demonstrated that DMC calculations can recover from the use of a HF trial wave function that has collapsed onto a discretized continuum solution, although larger electron binding energies are obtained when using a trial wave function for the anion that provides a more realistic description of the charge distribution, and, hence, of the nodal surface. For the geometry at which the cluster has a non-valence correlation-bound anion, both the inclusion of triples in the EOM-CC method and the inclusion of supplemental diffuse d functions in the basis set are important. DMC calculations with suitable trial wave functions give EBE values in good agreement with our best estimate EOM-CC result. AFQMC using a trial wave function for the anion with a realistic electron density gives a value of the EBE nearly identical to the EOM-CC result when using the same basis set. For the geometry at which the anion is bound in the HF approximation, the inclusion of triple excitations in the EOM-CC calculations is much less important. The best estimate EOM-CC EBE value is in good agreement with the results of DMC calculations with appropriate trial wave functions.
\end{abstract}

\maketitle

\section{\label{sec:Introduction}Introduction}

In recent years, there has been growing interest in a class of anions known as non-valence correlation-bound (NVCB) anions in which long-range correlation effects are crucial for the binding of the excess electron. \cite{Arai_H2O4_1,Arai_H2O4_2,Arai_NaCl, verlet_1,verlet_2,verlet_3,verlet_4,voora_c60_1,voora_c60_2,voora_c6f6,taehoon_c60,voora_pah,sommerfeld_nacl,cederbaum_2011}
By definition, NVCB anions are unbound in the Hartree-Fock (HF) approximation.
Due to their highly spatially extended charge distributions, large, flexible basis sets are required for the theoretical characterization of NVCB anions.
However, with such basis sets, the wave function from Hartree-Fock (HF) calculations on the excess electron system collapses onto the neutral plus an electron in an orbital that can be viewed as a discretized representation of a continuum solution.\cite{Arai_H2O4_1}
Methods that start from the HF wave function including second-order M{\o}ller-Plesset perturbation theory (MP2)\cite{MP2} or coupled-cluster singles and doubles with perturbative triples (CCSD(T))\cite{CCSDpT} do not recover from this collapse onto the continuum, while methods such as orbital-optimized MP2 (OOMP2)\cite{OOMP2} or Bruckner coupled-cluster\cite{BCC} can overcome this problem.\cite{Arai_H2O4_1}
The majority of calculations of NVCB anions have employed the equation-of-motion coupled-cluster singles and doubles (EOM-CCSD) method.\cite{EOMCCSD}
Among the NVCB anions studied computationally to date are \ce{C60}, \ce{C6F6}, TCNE, \ce{(NaCl)2}, \ce{Xe_{n}} clusters, large polyaromatic hydrocarbons, and certain \ce{(H2O)_{n}} clusters.\cite{Arai_H2O4_1,Arai_H2O4_2,Arai_NaCl,voora_c60_1,voora_c60_2,voora_c6f6,taehoon_c60,voora_pah,sommerfeld_nacl,cederbaum_2011}

The EOM-CCSD method displays an $\mathcal{O}(N^6)$ scaling with system size, and higher order EOM-CC methods are even more computationally demanding.
As a result, most of the calculations of NVCB anions carried out to date have not been fully converged with respect to basis set or the level of excitations treated in the EOM procedure.
We note, however, that by using domain-based local pair natural orbitals (DLPNO), electron affinity EOM-CCSD calculations have recently been carried out on systems described by up to 4,500 basis functions.\cite{dutta_dlpno_ea_eom_ccsd}

In the present work, we apply two quantum Monte Carlo (QMC) methods to the problem of calculating the electron binding energy (EBE) of the non-valence anion of a model \ce{(H2O)_4} cluster.
The first approach considered is fixed-node diffusion Monte Carlo (DMC),\cite{grimm_monte-carlo_1971,anderson_randomwalk_1975,anderson_quantum_1976,foulkes_quantum_2001} using various single Slater determinant (SD) and multideterminant (MD) trial wave functions.
DMC is a real-space method, with the major sources of error resulting from the use of finite time steps and the fixed-node approximation. 
The finite time step error can be largely eliminated by running calculations at different time steps and then extrapolating to the zero time step limit.
The fixed-node error results from imposition of a nodal surface via a trial wave function, which is necessary to ensure Fermionic behavior, and can be addressed by a variety of means including expanding the number of Slater determinants in the trial wave function or by applying the backflow transformation.\cite{backflow}
It is important to note that, by virtue of working in real space, fixed-node DMC energies are much less sensitive to the choice of the atomic basis set than methods such as EOM-CCSD that operate in a space of Slater determinants.

The second QMC approach considered is the auxiliary field QMC (AFQMC) method.\cite{AFQMC_1, AFQMC_2, AFQMC_3, zhang_quantum_2003, zhang_constrained_1997,motta2018ab,zhang2020ab}
AFQMC calculations sample an over-complete space of nonorthogonal Slater determinants.
The finite time step error can be mitigated as in DMC.  The error that arises from constraining the phase of the wave function to zero can be systematically reduced by improving the trial wave function. 
Phaseless AFQMC is additionally subject to the limitations of the atomic basis set employed.
DMC scales as $\sim\mathcal{O}(N^3)$ with system size, while AFQMC displays an $\sim\mathcal{O}(N^4)$ scaling in most implementations. 
One of the goals of these calculations is to determine whether DMC calculations can recover from the use of a trial wave function that has collapsed onto a discretized continuum orbital in the case of the excess electron.
Additionally, we explore whether correlation effects that are missing in EOM-CCSD are important for electron binding.

\begin{figure}
    \includegraphics[width=\columnwidth,keepaspectratio]{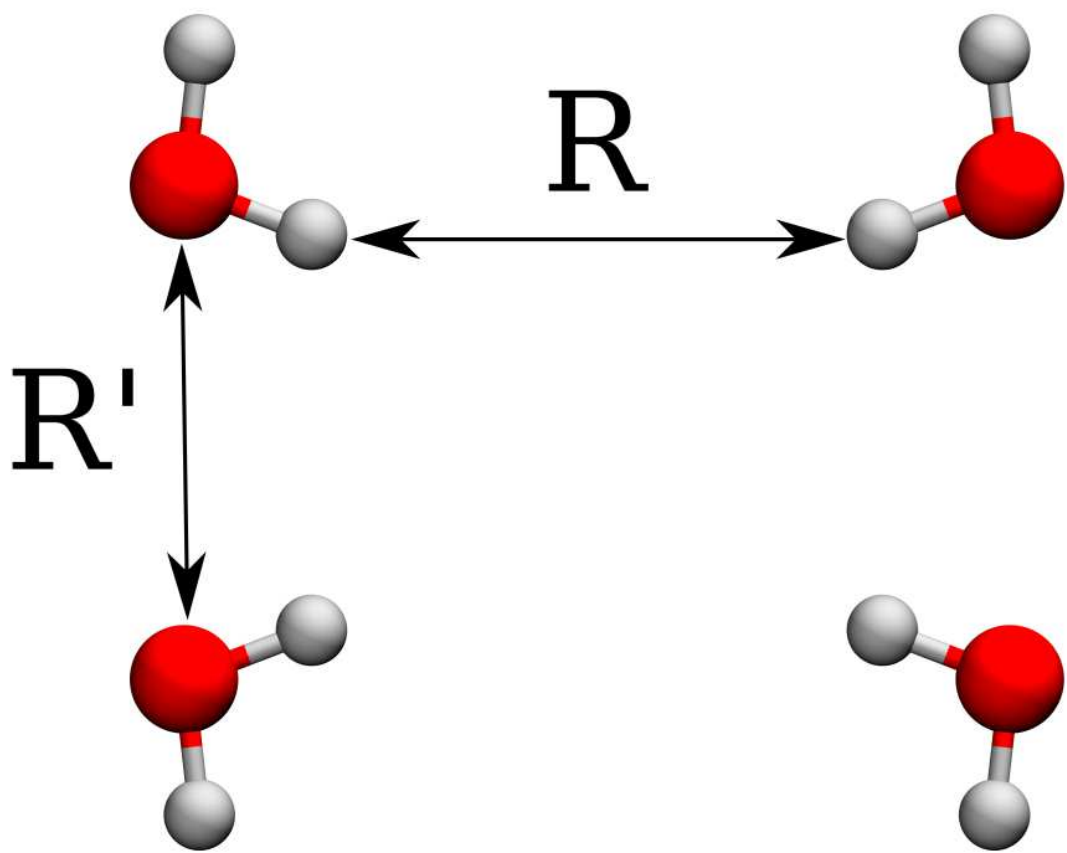}
    \caption{\label{fig:water4cluster} The  model \ce{(H2O)_4} system considered in this study. R$^{'}$ held fixed at \SI{3.46105}{\angstrom}, and R is either \SI{4}{\angstrom} or \SI{7}{\angstrom}. Image generated using VMD.\cite{VMD}}
\end{figure}

In our calculations, we employ a model \ce{(H2O)4} cluster that has been investigated in earlier studies by our group.\cite{Arai_H2O4_1,Arai_H2O4_2}
In this model, depicted in Figure~\ref{fig:water4cluster}, the monomers are arranged so that the net dipole moment is zero.
If the distance R is varied, with all other geometrical parameters held fixed, the system can be tuned from a regime (large R) that the excess electron weakly binds in the HF approximation to one (small R) at which it is not bound in the HF approximation. i.e., at which it is NVCB in nature.

\section{Methodology}
\subsection{EOM Coupled Cluster}
\label{sub:EOM}
The EOM methods considered in this study are EOM-MP2,\cite{EOMMP2} EOM-CCSD,\cite{EOMCCSD} EOM-CCSD(T)(a)$^*$,\cite{EOMCCSDpTASTAR} and EOM-CCSDT,\cite{EOMCCSDT1,EOMCCSDT2} listed in order of increasing sophistication in terms of treatment of correlation effects. 
In the EOM-MP2 and EOM-CCSD methods, the neutral molecule is treated at the MP2 and CCSD levels, respectively, and the amplitudes from these calculations are used to perform unitary transformation of the Hamiltonian.
This "dressed" Hamiltonian is then used to carry out a 1-particle plus 2-particle-1-hole configuration interaction (CI) calculation on the anion.
In the EOM-CCSDT method, the neutral species is first treated at the CCSDT level, and the transformed Hamiltonian is used to do CI calculation on the anion that includes up to 3-particle-2-hole configurations.
The EOM-CCSD(T)(a)$^*$ method includes in an approximate manner both triple excitations in the ground state coupled cluster calculations and 3-particle-2-hole excitations in the treatment of the anion.\cite{EOMCCSDpTASTAR}

The main basis set used for the EOM calculations reported in this study is aug-cc-pVTZ+7s7p, formed by supplementing the aug-cc-pVTZ Gaussian-type orbital (GTO) basis set\cite{cc-pVDTQZ,aug_for_cc-pVDTQZ} with a 7s7p set of diffuse functions centered at the middle of the cluster and similar to the set from Ref.~\onlinecite{Arai_H2O4_1}.
The exponents of the supplemental functions start at 0.023622, with each successive exponent being smaller by a factor of 3.2. 
However, as seen from Table~\ref{tab:diffusebasistuning}, the supplemental 7s7p set of diffuse functions can be truncated to 3s1p without significantly impacting the EBE as calculated at the EOM-CCSD level.
Moreover, as shown in Table~\ref{tab:corebasistuning}, expanding the main basis set (i.e., the non-supplemented portion) from aug-cc-pVTZ to aug-cc-pVQZ\cite{cc-pVDTQZ,aug_for_cc-pVDTQZ} makes only a small impact on the EBE (~4\% at R = \SI{4}{\angstrom}) .
In contrast, reducing the main basis set to aug-cc-pVDZ\cite{cc-pVDTQZ,aug_for_cc-pVDTQZ} leads to a ~14\% reduction in the EBE.
(These results were obtained using the EOM-MP2 method, but as seen from comparison of the results in Tables~\ref{tab:diffusebasistuning} and \ref{tab:corebasistuning}, using the aug-cc-pVTZ+3s1p basis set in both cases, the EBEs from the calculations with the EOM-CCSD and EOM-MP2 methods agree to within 0.5 meV.)
The smaller aug-cc-pVDZ+3s1p basis will be used in the EOM-CCSDT calculations, which would have been computationally prohibitive with aug-cc-pVTZ+7s7p or aug-cc-pVTZ+3s1p basis sets.
Finally, EOM-CCSD(T)(a)* calculations were carried out with aug-cc-pVTZ+3s1p3d basis sets, where the exponents of the d functions match those of the s and p functions, to assess the importance of supplemental d functions on the EBEs.
The EOM calculations utilized the frozen core approximation and were carried out using the CFOUR program.\cite{cfour_1,cfour_2}
\begin{table*}
    \caption{\label{tab:diffusebasistuning}Dependence of the total energies and the EBE of the model \ce{(H2O)4} cluster at R = \SI{4}{\angstrom} on the supplemental diffuse basis functions. Results obtained using the EOM-CCSD method.}
\begin{ruledtabular}
\begin{tabular}{l c c c}
basis set            & neutral (Ha)  & anion (Ha)    & EBE (meV)            \\
\hline
aug-cc-pVTZ      & -305.327947          & -305.331344          &   92.4            \\
aug-cc-pVTZ+1s   & -305.327953          & -305.332359          &  119.9             \\
aug-cc-pVTZ+2s   & -305.327957          & -305.334226          &  170.6              \\
aug-cc-pVTZ+3s   & -305.327958          & -305.334460          &  176.9              \\
aug-cc-pVTZ+7s   & -305.327958          & -305.334462          &  177.0              \\ \hline
aug-cc-pVTZ+7s1p & -305.327979          & -305.334604          &  180.3              \\
aug-cc-pVTZ+7s7p & -305.327987          & -305.334622          &  180.6              \\ \hline
aug-cc-pVTZ+3s1p & -305.327979          & -305.334602          &  180.2     
\end{tabular}
\end{ruledtabular}
\end{table*}

\begin{table}
    \caption{\label{tab:corebasistuning}Sensitivity of the EBE of the \ce{(H2O)4} model to the ``core'' basis set. Results obtained using the EOM-MP2 method.}
\begin{ruledtabular}
    \begin{tabular}{lccc}
                 & Neutral (Ha)          & Anion (Ha)             & EBE (meV)                      \\ \hline
\multicolumn{4}{c}{R = \SI{4.0}{\angstrom}}                                                        \\
aug-cc-pVDZ+3s1p & -305.0371957          & -305.0428558           & 154.0                     \\
aug-cc-pVTZ+3s1p & -305.3092869          & -305.3159306           & 180.8                     \\
aug-cc-pVQZ+3s1p & -305.4008845          & -305.4078074           & 188.4                     \\ \hline
\multicolumn{4}{c}{R = \SI{7.0}{\angstrom}}                                                   \\
aug-cc-pVDZ+3s1p & -305.0383747          & -305.0432259           & 132.0                     \\
aug-cc-pVTZ+3s1p & -305.3104923          & -305.3157472           & 143.0                     \\
aug-cc-pVQZ+3s1p & -305.4021640          & -305.4075716           & 147.1                    
\end{tabular}
\end{ruledtabular}
\end{table}

\subsection{DMC}
The DMC calculations were carried out using trial wave functions represented as products of one or more Slater determinants with a Jastrow factor with one-, two-, and three-body terms.\cite{Jastrow,jastrow_form_casino,QMCPACK_1} 
The parameters in the Jastrow factors were optimized using variational Monte Carlo (VMC), and the resulting trial wave functions were then employed in subsequent DMC calculations.
Three types of SD trial wave functions were employed. 
These used HF orbitals, Becke-Lee-Yang-Parr (B3LYP) DFT orbitals,\cite{B3,LYP,VWN, B3LYP_assembly} and natural orbitals (NOs) from small restricted single plus double excitation configuration interaction (SDCI) calculations designed to bind the excess electron when it is not bound in the HF approximation.
In addition, DMC calculation were carried out using MD trial wave functions, with the determinants being determined either from the restricted SDCI procedure or from configuration interaction using a perturbative selection made iteratively (CIPSI) calculations.\cite{CIPSI}
Details on these calculations are provided below.

To reduce the computational cost of the DMC calculations, the ccECP pseudopotentials\cite{ccecp_1,ccecp_2} were employed together with GTO basis sets that we designate as cc-pVDZ / ccECP, aug-cc-pVDZ / ccECP, aug-cc-pVDZ / ccECP+3s1p, and aug-cc-pVDZ / ccECP+7s7p. 
The "core" cc-pVDZ / ccECP\cite{ccecp_1,ccecp_2} basis set was designed for use with the ccECP pseudopotentials; the "aug" indicates that the diffuse aug functions from the aug-cc-pVDZ basis sets of Dunning and co-workers are included; and the 7s7p set of diffuse functions are those described above in the Section \ref{sub:EOM}.\cite{aug_for_cc-pVDTQZ}
The T-moves scheme was used to control the localization error for nonlocal pseudopotentials.\cite{casula_beyond_2006}

The double-zeta rather than the larger triple-zeta basis set was used as the core basis set due to the relative insensitivity of DMC calculations to the choice of the atomic basis set. 
For most of the DMC calculations a fixed population of 16,000 walkers and time steps of 0.001, 0.003, and 0.005 a.u. were employed, with the reported results obtained by linear extrapolation to zero time step.
However, this population is much larger and the time steps much smaller than what is actually required to achieve well converged energies with minimized finite time step and fixed population errors.
Indeed, DMC calculations using Hartree-Fock trial wave functions, larger time steps (specifically 0.05, 0.1, and 0.2 a.u.) and a smaller population of only 1,000 walkers produce an electron binding energy within error bars of that obtained using the smaller time steps and larger populations.
Additionally, a DMC calculation with a B3LYP trial wave function with a time step of 0.05 is in agreement with the values obtained with the smaller time steps and larger populations suggesting that these parameters do not depend strongly on the choice of starting orbitals.
In light of this, the 0.05 a.u. time step and smaller walker population were employed in the DMC calculations using CIPSI trial wave functions to mitigate the additional cost associated with the MD space. 
The VMC and DMC calculations were carried out using the QMCPACK code.\cite{QMCPACK_1,QMCPACK_2}
The orbitals for the SD-based trial wave functions and the restricted SDCI MD wave function were both generated using GAMESS,\cite{gamess_1,gamess_2,gamess_3} whereas the CIPSI wave functions were generated using the Quantum Package 2.0 code.\cite{QP2} 

\subsection{Restricted CI and CIPSI-generated Trial Wave Functions for DMC Calculations}
\label{subsec:rSDCI}
The restricted SDCI procedure employed the HF wave function for the neutral molecule and a specially tailored SDCI wave function for the anion, which included all symmetry-allowed single and double excitations, with the latter restricted so that one of the electrons excited is from the orbital occupied by the excess electron in the HF wave function.
This approach, when used with a flexible basis, gives a bound anion.
NOs were generated from the SDCI wave function of the anion and were used in a SD trial wave function for subsequent DMC calculations.
In addition, the SDCI wave function itself (expanded in terms of HF orbitals) was used in MD DMC calculations on the anion for R = \SI{4}{\angstrom}. 
In this case, a threshold of 0.001 on the magnitude of coefficients in the CI expansion was used in choosing the retained determinants.
This resulted in a wave function with 1,392 Slater determinants.

By design, the restricted SDCI wave function does not allow for change of the correlation energy of the valence electrons due to the presence of the excess electron.
This possibility is allowed for in the CIPSI MD trial wave functions.
The CIPSI calculations were carried out using B3LYP orbitals rather than Hartree-Fock orbitals because the former avoids the problem of collapse onto a discretized continuum solution at R =  \SI{4}{\angstrom}.\cite{B3,LYP,VWN}
Since the CIPSI calculations have not approached the full configuration interaction limit as indicated by the second-order perturbative correction to the energy, a judicious choice of starting orbitals is required to construct a physically meaningful trial wave function.
In order to generate compact wave functions for both the anion and the neutral, NOs were iteratively refined through successive CIPSI calculations, each beginning from a single determinant reference of natural orbitals from the previous iteration.
For each NO-generating CIPSI calculation, approximately 100,000 determinants were retained and used to generate NOs for the next iteration, for a total of six NO generation cycles. 
With the determinant of resulting NOs as a reference, a final CIPSI calculation was carried out, stopping when at least 150,000 determinants were included in the variational space for the anion and at least 100,000 determinants for the neutral. 
The resulting determinant spaces were used as the DMC trial wave functions.

Both the restricted SDCI and the CIPSI calculations used to generate the trial wave functions for subsequent DMC calculations were carried out using the ccECP pseudopotentials.
The aug-cc-pVDZ/ccECP+7s7p and aug-cc-pVDZ/ccECP+3s1p basis sets were used for the SDCI and CIPSI calculations, respectively.

\subsection{AFQMC}

AFQMC\cite{AFQMC_1, AFQMC_2, AFQMC_3, zhang_quantum_2003, zhang_constrained_1997,motta2018ab,zhang2020ab} utilizes the Hubbard-Stratonovich transformation \cite{hubbard1959calculation} to represent the imaginary-time propagator as a multi-dimensional integral over auxiliary-fields. 
Ground-state properties are sampled from a random walk in the space of non-orthogonal Slater determinants subject to the phaseless constraint\cite{zhang_quantum_2003} introducing a bias which can be systematically reduced based on the quality of the nodal surface of the trial wave function employed.
While sophisticated trial wave functions based on regularized orbital-optimized MP2 ($\kappa$-OOMP2)\cite{lee2020utilizing} or CASSCF\cite{rudshteyn2020predicting,shee2019achieving,kumar2020multiple} are required to obtain quantitative predictions for some biradicaloids and transition metals, high accuracy has been obtained, even for systems exhibiting non-trivial electron correlation such as dipole-bound anions,\cite{hao2018accurate} with single-determinant trial wave functions consisting of HF or Kohn-Sham orbitals.\cite{hao2018accurate,shee2019singlet}

In this work we perform calculations with a GPU implementation of AFQMC,\cite{shee2018phaseless}
utilizing single-precision floating-point arithmetic and two-electron integrals decomposed via a modified Cholesky decomposition (10$^{-5}$ cutoff).\cite{purwanto2011assessing} 
These calculations made use of the aug-cc-pVTZ+7s7p basis set, a small imaginary-time step of 0.005 a.u, and correlated all electrons.
For the neutral species and electrostatically bound anion (R = \SI{7}{\angstrom}), the Hartree-Fock wave function was used as the trial wave function.
For the NVCB anionic species (R = \SI{4}{\angstrom}), a SD trial wave function comprised of natural orbitals from the restricted SDCI calculation as detailed in Section~\ref{subsec:rSDCI} (but now carried out without pseudopotentials) was used.

\subsection{Radial orbital densities}
To compare the description of the charge distribution of the excess electron as calculated using different theoretical methods, we generate radial electron density plots.
This choice is motivated by the fact that the excess electron occupies an orbital belonging to the totally symmetric representation.
The radial electron densities are generated by numerically integrating over the angular components of the singly occupied molecular or natural orbital.
First, Molden files are created from the output data from the various generating programs using cclib when supported.\cite{cclib}
With the Molden files as input, PySCF is used to generate the electron density on a uniform radial grid and 5810 point Lebedev-Laikov angular grid as tabulated in quadpy.\cite{pyscf1,pyscf2,lebedevlaikov,quadpy}
Finally, a numerical integration is performed over the angular components. 
An example of this workflow is presented in detail in the Supplementary Information.

\section{Results}
The EBEs obtained from the EOM and AFQMC calculations are summarized in Table~\ref{tab:EOM}, and the results from the various DMC calculations are summarized in Table~\ref{tab:DMC}.
We consider first the results obtained for R = \SI{4}{\angstrom}, for which HF calculations do not bind the excess electron. 

\subsection{Results for R = \SI{4}{\angstrom}: the correlation bound region}
From Table~\ref{tab:EOM}, it is seen that the EOM-CCSD/aug-cc-pVTZ+7s7p calculations give a value of the EBE of 181 meV for the \ce{(H2O)4} cluster model at R = \SI{4}{\angstrom}.
This increases to 196 meV with the EOM-CCSD(T)(a)$^*$ method.
The AFQMC calculations using the same basis set and for the anion a single determinant of NOs from the restricted SDCI calculation for the trial wave function produce an EBE value of 194 $\pm$ 10 meV, comparable to the EOM-CCSD(T)(a)$^*$ result. 
The EOM-CCSD(T)(a)$^*$ and EOM-CCSDT EBE values calculated with this basis set are nearly identical, demonstrating that the approximate treatment of triples in the former procedure introduces a negligible error in the EBE.
The contribution of supplemental diffuse functions was checked using the EOM-CCSD(T)(a)$^*$ method and the aug-cc-pVTZ+3s1p2d basis set.
These calculations reveal that the inclusion of the supplemental diffuse d functions leads to a $\sim$10 meV increase in the EBE. With the inclusion of this correction, we obtain an estimated EOM-CCSDT EBE of 212 meV. It is expected that the inclusion of the supplemental d functions in the basis set used for the AFQMC calculations would lead to a similar increase in the EBE obtained using that method.

The restricted SDCI procedure, by itself, is not expected to give an accurate value of the EBE and is designed to generate appropriate trial wave functions for DMC or AFQMC calculations on the anion.
In fact, the EBE resulting from the HF treatment of the neutral and the restricted SDCI treatment of the anion using the aug-cc-pVTZ+7s7p basis set is 345 meV, appreciably larger than the EOM and AFQMC values.
This over-binding is due in part to the fact that the restricted SDCI wave function, like the HF wave function, overestimates the magnitude of the dipole moment of the water molecules, resulting in a too favorable electrostatic interaction.
We also constructed a single determinant trial wave function for the anion using the natural orbitals of the restricted SDCI expansion.
We note also that the single determinant of NOs generated from the restricted SDCI wave function and using the aug-cc-pVTZ+7s7p basis set places the anion 160 meV above the neutral when the latter is treated in the HF approximation. 
This is not surprising since this calculation neglects correlation effects other than those incorporated in the determination of the orbitals.
What is important is that the approaches based on the restricted SDCI procedure provide a realistic description of the orbital occupied by the excess electron and avoid the collapse onto the discretized continuum as was observed with the HOMO in the HF calculations. 

\begin{table}
    \caption{\label{tab:EOM} EBEs of the \ce{(H2O)4} model calculated using HF, EOM, and AFQMC methods and employing the aug-cc-pVTZ+7s7p basis set.}
\begin{ruledtabular}
\begin{tabular}{lr}
Method & EBE (meV)                               \\
\hline
\multicolumn{2}{c}{R = \SI{4.0}{\angstrom}}     \\ 
HF            & -0.4        \\
EOM-CCSD      & 180.6          \\
EOM-CCSD(T)(a)$^*$ & 195.8          \\
EOM-CCSDT     & 197.5\footnote{\label{eomccsdt}Estimated by adding the difference of EBEs from the EOM-CCSD(T)(a)$^*$ and EOM-CCSDT calculations with the aug-cc-pVDZ+3s1p basis set to the value from EOM-CCSD(T)(a)$^*$/aug-cc-pVTZ+7s7p.} 
(212.0)
\footnote{\label{dfunc}Estimated by adding the difference between the EBEs calculated with the EOM-CCSD(T)(a)$^*$ with the aug-cc-pvTZ+3s1p and aug-cc-pVTZ+3s1p3d basis sets to the estimate described in footnote \ref{eomccsdt} to assess the effect of incorporating diffuse d functions into the basis.}  \\
AFQMC SD/HF(N)//SD/NO SDCI(A)     & 194 $\pm$ 10 \\ \hline
\multicolumn{2}{c}{R = \SI{7.0}{\angstrom}}                    \\ 
HF            & 41.3        \\
EOM-CCSD      & 140.2    \\
EOM-CCSD(T)(a)$^*$ & 141.7    \\
EOM-CCSDT    & 143.3\textsuperscript{\ref{eomccsdt}}  (154.2)\textsuperscript{\ref{dfunc}}     \\
AFQMC SD/HF     & 181 $\pm$ 5 \\ 
\end{tabular}
\end{ruledtabular}
\end{table}

In light of the close agreement between the EOM-CCSD(T)(a)$^*$ and AFQMC values of the EBE of the \ce{(H2O)4} model at R = \SI{4}{\angstrom}, when using a comparable basis sets in the two approaches it is relevant to determine whether DMC calculations with sufficiently flexible trial wave functions give an EBE close to the AFQMC and EOM values consistent with these results.
DMC calculations using HF trial wave functions together with the aug-cc-pVDZ/ccECP+7s7p basis set give an EBE of 183 $\pm$ 10 meV, appreciably smaller than the EOM-CCSD(T)(a)$^*$ and AFQMC values.
Interestingly, essentially the same EBE is obtained from the DMC calculations using a Slater determinant of HF orbitals expanded in the aug-cc-pVDZ/ccECP basis set without the 7s7p supplemental set of diffuse functions.
However, if the aug diffuse functions are also removed, the DMC calculations fail to bind the excess electron.
We believe that this is a consequence of the fact that with the cc-pVDZ basis set there is a near zero probability of sampling regions of space at large distances from the molecule, which are important for describing the charge distribution of the excess electron.
\begin{table}
    \caption{\label{tab:DMC} EBEs of the \ce{(H2O)4} model calculated using the DMC method and various trial wave functions.\protect\footnote{SD/X indicates that the trial wave function employed a single Slater determinant with X (either HF or B3LYP) orbitals. When different types of trial wave functions are used for the neutral (N) and anion (A) this is indicated by the double slash.}}
\begin{ruledtabular}
    \begin{tabular}{llr}
wave function                  & basis set        &  EBE (meV) \\ \hline
\multicolumn{3}{c}{R = \SI{4.0}{\angstrom}}               \\ 
SD/HF                         & aug-cc-pVDZ+7s7p &  183 $\pm$ 10 \\
SD/HF                         & aug-cc-pVDZ      &  176 $\pm$ 12 \\
SD/HF                         & cc-pVDZ          & -528 $\pm$ 25 \\ 
SD/B3LYP                      & aug-cc-pVDZ+7s7p &  212 $\pm$ 11 \\
SD/HF(N)//SD/NO SDCI(A)    & aug-cc-pVDZ+7s7p &  205 $\pm$ 10 \\
SD/HF(N)//MD/NO SDCI(A)    & aug-cc-pVDZ+7s7p &  202 $\pm$ 12 \\
MD/CIPSI NO                   & aug-cc-pVDZ+3s1p &  190 $\pm$ 9  \\ \hline
\multicolumn{3}{c}{R = \SI{7.0}{\angstrom}}                    \\ 
SD/HF                         & aug-cc-pVDZ+7s7p &   141 $\pm$ 14 \\ 
SD/B3LYP                      & aug-cc-pVDZ+7s7p &   164 $\pm$ 9 \\
SD/HF(N)//SD/NO SDCI(A)    & aug-cc-pVDZ+7s7p &  160 $\pm$ 9 \\
MD/CIPSI NO                   & aug-cc-pVDZ+3s1p &    159 $\pm$ 8 \\
\end{tabular}
\end{ruledtabular}
\end{table}

A significantly larger value of the EBE is obtained from SD DMC calculations using B3LYP orbitals in place of HF orbitals.
The resulting EBE of 212 $\pm$ 11 meV, within statistical error, agrees with the EOM-CCSD(T)(a)$^{*}$ and AFQMC values.
A similar value of the EBE is obtained from DMC calculations using a single determinant of HF orbitals for the neutral cluster and a single determinant of natural orbitals from the restricted SDCI procedure described in Section~\ref{subsec:rSDCI} for the anion.
DMC calculations using a SD of HF orbitals for trial wave function of the neutral and a trial wave function for the anion retaining 1,392 of the most important determinants from the restricted SDCI calculation gives an EBE of 202 $\pm$ 12 meV, close to the values obtained using the single determinants B3LYP orbitals or of NOs from the SDCI calculation (for the anion). 
The DMC value of the EBE resulting from the anionic trial wave function using a SD of NOs from the restricted SDCI MD calculation results is 205 $\pm$ 10meV, similar to that from DMC calculations using as trial wave functions the MD restricted SDCI wave function for the anion and the HF wave function for the neutral.

\begin{figure}
    \caption{\label{fig:orbitalsR4} Radially integrated charge densities of the singly occupied orbitals from HF and B3LYP calculations and the singly occupied natural orbital from EOM-CCSD, SDCI, and CIPSI calculations  of the model (\ce{(H2O)4}) cluster anion at R = \SI{4}{\angstrom}. All plots generated using Matplotlib.\cite{matplotlib}}.
    \includegraphics[width=\columnwidth,keepaspectratio]{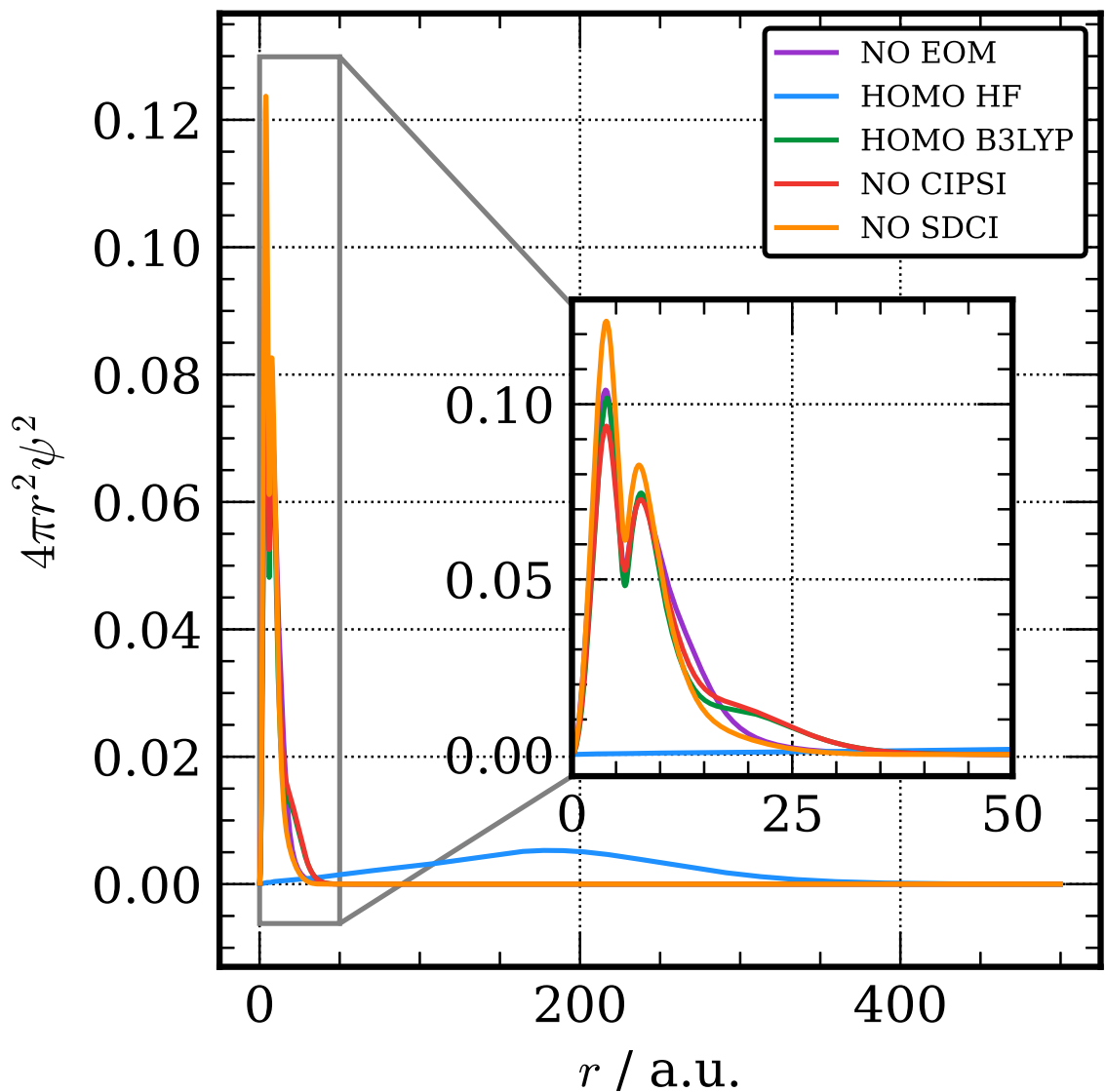}
\end{figure}

Figure~\ref{fig:orbitalsR4} compares the radial charge distributions of the singly occupied orbital from the HF and B3LYP calculations on the excess electron system as well as of the NOs associated with the excess electron from EOM-CCSD, restricted SDCI and CIPSI calculations. 
The collapse of the singly occupied orbital from the HF calculations onto a discretized continuum orbital is readily apparent. 
In contrast, the NOs from the EOM-CCSD and restricted SDCI calculations and the singly occupied orbital from the B3LYP calculation on the anion are more localized and are qualitatively similar to one another. 
These results are consistent with the nodal surface for the anion being significantly improved when using a SD trial wave function that has a physically reasonable charge distribution for the orbital occupied by the excess electron.
Thus, although DMC calculations do recover from the collapse of the HF trial wave function onto a discretized continuum solution in the case of the anion, starting with such a trial function leads to a greater nodal surface error for the anion than for the neutral cluster.
However, we also note that the radial distribution function of the singly occupied orbital from the B3LYP calculation on the anion has a spurious peak near 25 atomic units from the center of the cluster.
This is likely a consequence of the self-interaction error in the B3LYP functional.
The relevant NO extracted from the CIPSI calculations, which were carried using B3LYP orbitals, exhibits a similar shoulder.

Our final set of DMC calculations at R = \SI{4}{\angstrom} used MD trial wave functions determined from CIPSI calculations for the neutral and anionic clusters. 
The strategy used in performing the CIPSI calculations was presented in Section \ref{subsec:rSDCI}, where it was noted that these calculations, unlike those with the restricted SDCI wave functions, allow for the correlation between the valence electrons change due to the the presence of the excess electron. 
The DMC calculations using the CIPSI trial wave function resulted 190 $\pm$ 9 meV for R = \SI{4}{\angstrom}, slightly under-binding compared to the single determinant DMC value of the EBE obtained using B3LYP orbitals though in close agreement with the results of DMC calculations carried out with the restricted SDCI trial wave function.

\subsection{Results for R = \SI{7}{\angstrom}: the electrostatically bound region}

We now consider the results obtained for the \ce{(H2O)4} cluster model at R = \SI{7}{\angstrom}, for which HF calculations with the aug-cc-pVTZ+7s7p basis set bind the excess electron by 41 meV.
In this case, the EOM-CCSD and EOM-CCSD(T)(a)$^{*}$ calculations give EBEs of 140 meV and 142 meV, respectively.
Thus unlike the situation for R = \SI{4}{\angstrom}, the inclusion of triples in the EOM-CC procedure is relatively unimportant at R = \SI{7}{\angstrom}.
The DMC calculations using SD HF trial wave functions give an EBE of 141 $\pm$ 14 meV, while the DMC calculations using as trial wave functions single determinants of B3LYP orbitals, single determinants generated using the restricted SDCI procedure, or MD trial wavefunctions generated using the CIPSI procedure give similar EBEs values ranging  from 159 $\pm$ 8 to 164 $\pm$ 9 meV. 

Since the anion is bound in the HF approximation at R = \SI{7}{\angstrom}, we also were able to calculate EBEs using separate, frozen-core coupled-cluster calculations for the neutral and anion with the following coupled-cluster methods: coupled-cluster singles, doubles, and a perturbative treatment of triples $\Delta$CCSD(T)\cite{CCSDpT}, coupled-cluster singles, doubles, and triples ($\Delta$CCSDT)\cite{deltaccsd,CCSDT1,CCSDT2,CCSDT3}, and CCSDT with the perturbative treatment of quadruple excitations ($\Delta$CCSDT(Q)).\cite{CCSDTpQ1} methods.
The $\Delta$ indicates that the EBE is derived from the energy difference between the separate calculations on the neutral and anion. The $\Delta$CCSDT and $\Delta$CCSDT(Q) calculations were carried out with only the aug-cc-pVDZ+3s1p basis set.
These calculations indicate that full treatment of the triples, and even approximate treatment of the quadruple excitation contributions, has less than a 1 meV effect on the EBE of the \ce{(H2O)4} cluster model at R = \SI{7.0}{\angstrom}.
On the other hand, the inclusion of diffuse d function in the supplemental set of functions leads to a 12 meV increase in the EBE. With this correction we obtain an estimated EOM-CCSDT EBE of 154 meV, which is in good agreement with the DMC results using suitable trial wave functions.

\begin{figure}
    \caption{\label{fig:orbitalsR7} Radially integrated charge densities of the singly occupied orbitals from HF and B3LYP calculations and the singly occupied natural orbital from EOM-CCSD, restricted SDCI, and CIPSI calculations  of the model (\ce{(H2O)4}) cluster anion at R = \SI{7}{\angstrom}.}
    \includegraphics[width=\columnwidth,keepaspectratio]{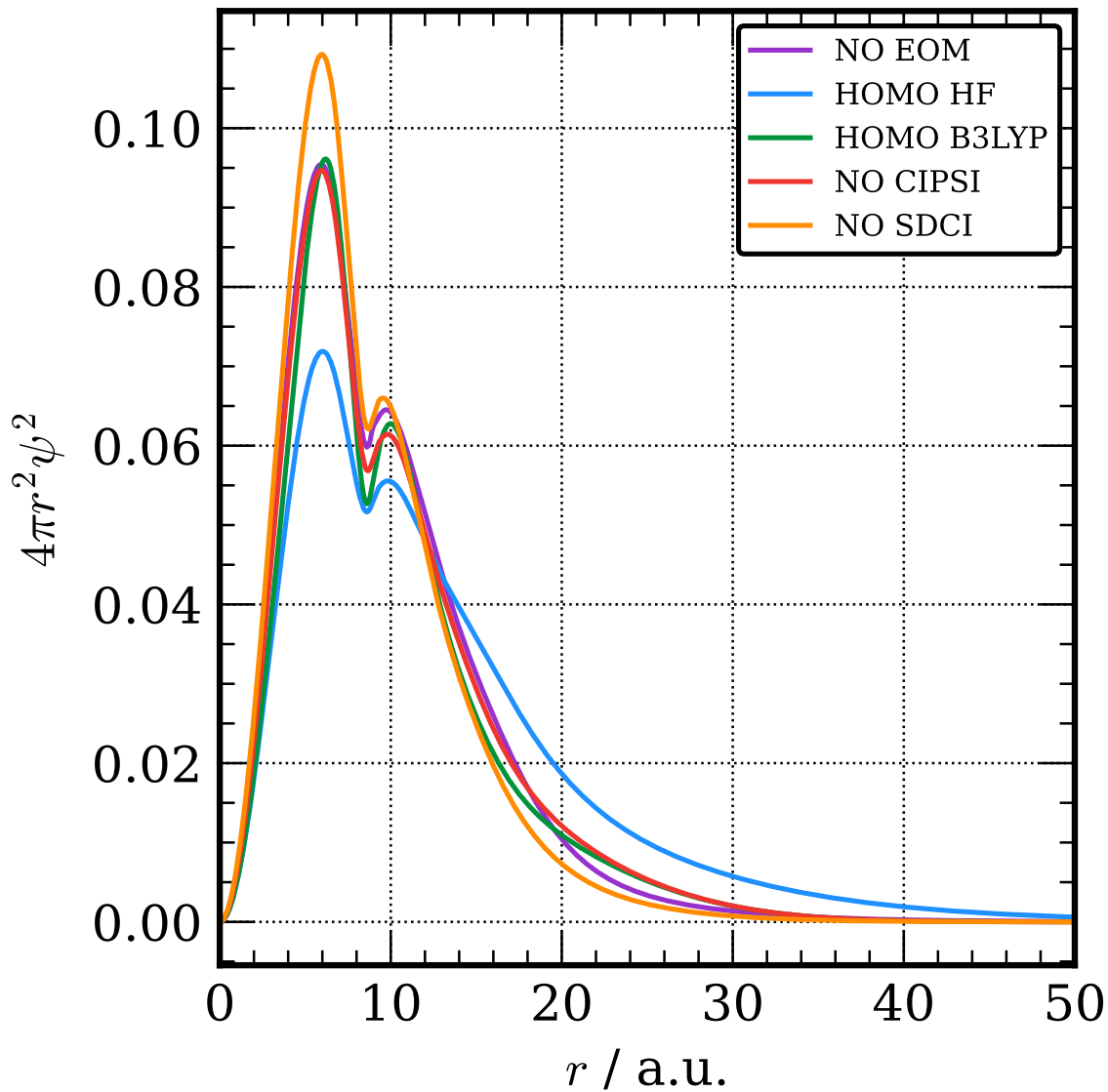}
\end{figure}

The AFQMC calculations give an EBE of 181 $\pm$ 5 meV, significantly larger than the EOM-CC results or DMC values.
This most likely reflects an inadequacy of the HF wave function used for the anion in the AFQMC calculations.
Support for this interpretation is provided by examination of Figure~\ref{fig:orbitalsR7}, which shows the radial charge distribution of the excess electron for the \ce{(H2O)4} model at R = \SI{7}{\angstrom}.
From this figure it is seen that although that the HF wave function has not collapsed onto the continuum as it did in the R = \SI{4}{\angstrom} cluster, it is still much more diffuse than that from calculations that include correlation effects. It is also seen from comparisons of Figures~\ref{fig:orbitalsR4} and \ref{fig:orbitalsR7} that the charge distribution associated with the NO occupied by the excess electron in the EOM-CCSD calculations for the cluster with R = \SI{7}{\angstrom}, is more radially extended than that at R = \SI{4}{\angstrom}.
Another noticeable difference between the charge density plots for R = \SI{7}{\angstrom} and \SI{4}{\angstrom} is the reduction of the long-range shoulder in the radial charge distribution of the HOMO from the B3LYP calculations on the anion and in the relevant NO from the CIPSI calculations on the anion carried out using B3LYP orbitals, suggesting that self-interaction errors are less problematical at R = \SI{7}{\angstrom}.

\section{\label{sec:Conclusion}Conclusion}
In this study we have applied various EOM-CC methods and two different quantum Monte Carlo methods to calculate the EBE of a model \ce{(H2O)4} cluster at two geometries, one at which the anion is bound in the HF approximation and the other at which it is not.
Diffusion Monte Carlo calculations using single determinant trial functions based on Hartree-Fock orbitals are shown to bind the excess electron even when the initial wave function for the anion has collapsed onto the neutral plus discretized continuum orbital. 
However, such calculations significantly underestimate the EBE, whereas SD DMC calculations using trial wave functions for the anion with a more realistic charge distribution for the excess electron give larger EBE values that are in close agreement with our best estimate EOM-CCSDT values for both geometries considered.

For R = \SI{4}{\angstrom}, at which the anion is correlation bound, use of such trial wave functions accurately reflecting the physical charge density resulted in AFQMC-predicted EBE values in agreement with the EOM-CCSD(T)(a)$^*$ result (when using comparable basis sets). 
However, at R = \SI{7}{\angstrom}, AFQMC calculations with HF trial wave functions significantly overestimate the EBE compared to EOM-CC and DMC values, suggesting the need for an improved trial wave functions in this case.
For the \ce{(H2O)4} model system, the restricted SDCI represents an economical way to create trial wave functions for QMC calculations on non-valence anions that are not bound in the Hartree-Fock approximation. 
However, it remains to be seen if this strategy will be as effective for systems in which the neutral species is more strongly correlated than the model \ce{(H2O)4} cluster.

Finally, we note that at R = \SI{4}{\angstrom}, for which the anion is NVCB in nature, the most frequently used method to characterize such anions, EOM-CCSD, underestimates the EBE by about 10\% compared to the result of EOM-CCSDT calculations. 
Both DMC and AFQMC are viable alternatives to high order EOM methods, and while more computationally demanding for the \ce{(H2O)4} cluster, they demonstrate lower scaling with system size than EOM methods, making them attractive for the characterization of non-valence anions of much larger systems.

\section*{Supplementary Material}
See Supplementary Materials for geometries, basis sets, additional simulation details, and instructions on generating radial orbital density plots.

\begin{acknowledgments}
We acknowledge valuable discussions with Shiwei Zhang.
This research was carried out with the support of grant CHE-1762337 from the National Science Foundation.
SU acknowledges fellowship support from the Pittsburgh Quantum Institute.
AD acknowledges funding support from National Science Foundation CHE-1807683.
Computational resources for the EOM, DMC, and CIPSI calculations were provided by the Center for Research Computing, University of Pittsburgh.
The AFQMC calculations used resources of the Oak Ridge Leadership Computing Facility at the Oak Ridge National Laboratory, which is supported by the Office of Science of the U.S. Department of Energy under contract no.~DE-AC05-00OR22725.
\end{acknowledgments}

\section*{AIP Publishing Data Sharing Policy}
The data that support the findings of this study are openly available in a public, version-controlled repository at \url{https://github.com/shivupa/Water4_JCP_Special_Issue_Supplemental_Material}.

\bibliography{qmc.bib}
\end{document}